\documentclass[sigconf,nonacm]{acmart}
\AtBeginDocument{%
  }

\copyrightyear{2026}
\acmYear{2026}
\setcopyright{rightsretained}
\acmConference[WWW '27]{The ACM Web Conference 2027}{May 10--14, 2027}{Dublin, Ireland}
\acmBooktitle{The ACM Web Conference 2027 (WWW '27), May 10--14, 2027, Dublin, Ireland}

\usepackage{algorithm}
\usepackage{algorithmic}

\usepackage{subfig}
\usepackage{placeins}
\begin{document}

\title{When the Mythical Man-Month Meets Agentic Web Systems: Structured Coordination for Scalable Development}

\author{Zhiyao Wu}\authornote{Equal contributions.}
\email{bc31072@um.edu.mo}
\orcid{0009-0005-0827-0720}
\affiliation{%
  \institution{University of Macau}
  \city{Macau}
  \country{China}
}

\author{Zi Liang}\authornotemark[1]
\email{zi1415926.liang@connect.polyu.hk}
\affiliation{%
  \institution{Hong Kong Polytechnic University}
  \city{Hong Kong}
  \country{China}}

\author{Haibo Hu}\authornote{Corresponding author.}
\email{haibo.hu@polyu.edu.hk}
\affiliation{%
  \institution{Hong Kong Polytechnic University, PolyU Research Centre for Privacy and Security Technologies in Future Smart Systems}
  \city{Hong Kong}
  \country{China}}

\renewcommand{\shortauthors}{Zhiyao Wu, Zi Liang, and Haibo Hu}

\begin{abstract}
As agentic systems are increasingly deployed to support complex Web tasks, multi-agent architectures offer greater parallelism, specialization, and task coverage. However, increasing the number of agents does not necessarily improve system performance, as communication, synchronization, redundant work, and coordination overhead may offset the benefits of additional agents. This presents a fundamental scaling challenge for agentic Web systems: how can agent populations be organized to balance task performance and coordination costs? To address this challenge, we systematically investigate the relationship between agent population size, per-agent token consumption, and coordination overhead in collaborative agentic systems. Experiments on complex Web-oriented software development tasks reveal a non-linear scaling pattern: per-agent token consumption initially decreases as the agent population grows, but eventually exhibits diminishing returns. Motivated by these findings, we introduce the Surgical Team Model (STM), a structured coordination framework that organizes agents through role specialization, constrained inter-agent communication, and coordinator-mediated task aggregation. Experiments on full-stack Web application development tasks show that STM achieves competitive task completion quality and comparable token consumption relative to representative baselines, while incurring higher execution latency.

\end{abstract}




\maketitle

\section{Introduction}
Large Language Models (LLMs) and autonomous agents are increasingly transforming how complex tasks on the Web are performed. Agentic systems are being integrated into Web-facing applications such as search, information retrieval, software development, content generation, and interactive services, where agents can autonomously plan, execute, and coordinate multiple subtasks~\cite{wang2023metagpt,du2023chatdev,jiang2023autogen,chen2023agentverse,jiang2024crewai}. Compared with single-agent systems, multi-agent architectures provide additional parallelism, specialization, and task coverage, making them promising for complex tasks that require diverse capabilities and coordinated execution. As multi-agent systems become increasingly sophisticated, however, their effectiveness depends not only on the capabilities of individual agents but also on how many agents are deployed and how they interact. As agentic capabilities become increasingly integrated into Web applications, agentic Web systems are emerging as a new form of Web application infrastructure, where scalable agent coordination becomes an important systems concern~\cite{Su2026,Dong2026}. This makes the scalability and coordination of agentic Web systems an important research challenge, directly related to the design and orchestration of AI agents and multi-agent systems on the Web.

A fundamental challenge therefore arises in scaling multi-agent systems: \emph{does adding more agents always improve system performance?} Recent studies have explored scaling LLM-based systems by increasing the number of agents~\cite{li2024moreagents}. However, simply expanding the agent population does not necessarily yield proportional performance gains~\cite{google2025scaling,yang2026understanding,tran2026singleagent,chen2026diversitycollapse}. As more agents participate in a task, communication, synchronization, redundant computation, and coordination overhead can increase. Consequently, additional agents may provide diminishing returns or even reduce overall efficiency. This phenomenon resembles the classical \emph{Mythical Man-Month} effect~\cite{brooks1995mythical} in software engineering, where increasing manpower can paradoxically reduce productivity. Understanding this scaling behavior is therefore important for designing scalable multi-agent systems with effective coordination.

Existing research on multi-agent systems has primarily focused on improving agent collaboration frameworks~\cite{du2023chatdev,wang2023metagpt,jiang2023autogen,chen2023agentverse,jiang2024crewai} and reducing communication costs through more efficient interaction strategies~\cite{zheng2025thought,zhu2024sparse}. These approaches demonstrate the benefits of parallel execution and structured collaboration, but the relationship between agent population and coordination overhead remains insufficiently characterized. In particular, existing systems generally do not explicitly examine how coordination overhead evolves as the agent population increases or identify the efficient operating region of different agent population sizes. As a result, the trade-off between parallelism and coordination remains insufficiently understood, limiting principled guidance for determining an effective agent population in scalable multi-agent systems.

To address this gap, we conduct a systematic empirical study of agent population scaling in collaborative multi-agent systems. Using complex Web-oriented software-development tasks, we examine how token consumption and per-agent efficiency change as the number of agents increases. Our experiments reveal a consistent non-linear scaling pattern: increasing the agent population initially improves per-agent efficiency, whereas larger populations introduce increasing coordination-related costs and eventually lead to diminishing returns. We further keep task specifications, role prompts, model configurations, and decoding parameters consistent across experimental settings to provide a controlled comparison of different agent population sizes. These findings provide an empirical basis for understanding the scalability limits of multi-agent collaboration and motivating structured coordination mechanisms.

Motivated by these observations, we introduce the Surgical Team Model (STM), a structured coordination mechanism designed to manage agent population and interaction overhead. Inspired by the organization of specialized teams, STM assigns agents to task-specific roles, constrains unnecessary inter-agent communication, and employs coordinator-mediated task aggregation. Rather than maximizing the number of agents, STM organizes agents through a structured interaction topology that preserves the benefits of parallelism and specialization while limiting coordination overhead. Experiments demonstrate that STM achieves competitive task completion quality with comparable token-level cost to representative baselines, while incurring higher execution latency. These findings suggest that structured coordination can improve the scalability of multi-agent systems by controlling coordination overhead.

In summary, this paper makes the following contributions:

\begin{itemize}

\item \textbf{Empirical characterization of agent population scaling.} We systematically investigate how agent population affects token-based efficiency in collaborative multi-agent systems, identifying a non-linear scaling pattern and characterizing the efficient operating regime under unstructured scaling.

\item \textbf{Structural analysis of coordination overhead.} We analyze how inter-agent interaction patterns contribute to coordination overhead in collaborative multi-agent systems, examining the structural cost implications of unstructured communication as agent populations grow.

\item \textbf{Surgical Team Model for structured coordination.}
Motivated by the observed scaling behavior, we introduce STM, a role-based coordination mechanism that organizes agents through specialized roles, constrained communication, and coordinator-mediated aggregation. STM improves the quality--cost--latency trade-off compared with existing multi-agent frameworks.
\end{itemize}

\section{Related Work}
\label{sec:related}

Prior work on multi-agent systems for software development, particularly in Web application development, can be broadly organized into two directions: agent scaling and coordination which examines communication mechanisms and the effects of increasing agent populations, and role-based collaboration which uses specialized agent roles to support task decomposition and coordination.

\subsection{Agent Scaling and Coordination}

Existing studies have explored communication and coordination mechanisms to improve the efficiency of multi-agent systems, including sparse message passing \cite{zhu2024sparse}, adaptive communication topologies \cite{li2023adaptive}, large-scale coordination strategies \cite{qian2025macnet}, and communication protocols \cite{zheng2025thought}. These approaches primarily aim to reduce unnecessary inter-agent communication and improve coordination efficiency.

Recent studies have also investigated how agent population affects system performance. Li et al.~\cite{li2024moreagents} examine the benefits of increasing the number of agents, while subsequent work shows that performance gains can saturate or decline as agent populations grow \cite{kim2026capable,yang2026understanding,tran2026singleagent,chen2026diversitycollapse}. Such limitations have been attributed to redundant outputs, limited diversity, and increasing coordination costs. However, existing studies largely examine communication efficiency or population scaling separately, with limited systematic analysis of the joint relationship among agent population, coordination overhead, and task performance in collaborative software-development settings. Our work addresses this gap through a controlled empirical study across a broad range of agent populations.

\subsection{Role-based collaboration}

Role-oriented frameworks such as MetaGPT \cite{wang2023metagpt}, ChatDev \cite{du2023chatdev}, AgentVerse \cite{chen2023agentverse}, and CrewAI \cite{jiang2024crewai} assign specialized roles to facilitate task decomposition and collaborative execution. AutoGen \cite{jiang2023autogen} and CodeAgents \cite{paranjape2024codeagents} further support structured multi-agent interaction for code generation and software development, while CAMEL \cite{li2023camel} explores role-playing based multi-agent collaboration.

These approaches demonstrate the benefits of role specialization and structured collaboration. However, their primary focus is on organizing agent capabilities and task interactions rather than explicitly managing agent population and coordination overhead as system scale increases. Our work studies this scaling problem and introduces the Surgical Team Model (STM), which combines role specialization with constrained communication and coordinator-mediated aggregation to improve scalability and efficiency.

\textbf{Summary:} \textbf{Summary:} Overall, existing studies have explored agent scaling, communication efficiency, and role-based collaboration, but relatively limited attention has been paid to how agent population, coordination overhead, and task performance jointly evolve as multi-agent systems scale. We address this gap through a systematic study of agent population scaling and introduce STM to enable structured coordination for scalable development in agentic Web systems.

\section{Empirical Scaling Behavior of Agent Population}
\label{optimal_size_identification}

This section investigates how the agent population affects the scalability and token-based efficiency of agentic Web systems in Web application development. We formulate an empirical scaling model to characterize changes in per-agent token consumption as the agent population increases, and evaluate the resulting scaling behavior across diverse Web application development tasks.

\subsection{Agent Population Scaling Model}
\label{sec:scaling_model}

Let $n$ denote the number of agents participating in a software development task. We characterize the token-based cost at population size $n$ using the total token consumption $T_t(n)$. To compare systems with different agent populations, we define the per-agent token consumption as
\begin{equation}
T_p(n) = \frac{T_t(n)}{n}.
\label{eq:per_agent_token}
\end{equation}
A lower value of $T_p(n)$ indicates lower average token consumption per engineering agent and serves as a proxy for token-based efficiency. This metric does not directly capture task quality or end-to-end execution latency, which are evaluated separately.

To characterize the effect of increasing the agent population, we define the marginal efficiency gain as
\begin{equation}
G(n) = T_p(n) - T_p(n+1).
\label{eq:marginal_gain}
\end{equation}
When $G(n)>0$, adding an agent reduces the per-agent token consumption and therefore improves token-based efficiency. When $G(n)<0$, adding an agent increases the average token consumption per agent and reduces token-based efficiency. When $G(n)$ approaches zero, the benefit of increasing the agent population becomes limited.

As the agent population grows, we characterize changes in marginal efficiency gains using the second-order difference
\begin{equation}
\Delta G(n) = G(n+1)-G(n).
\label{eq:second_order_gain}
\end{equation}
When $G(n)>0$ and $\Delta G(n)<0$, the marginal efficiency gain decreases as the agent population increases, indicating diminishing returns in token-based efficiency. When $G(n)\leq 0$, adding an agent no longer improves per-agent token efficiency. This analysis characterizes changes in marginal gains without assuming a specific functional form for the scaling behavior.

Based on the observed per-agent token consumption, we define the empirical optimal agent population as
\begin{equation}
n^* = \arg\min_{n \in N} T_p(n),
\label{eq:optimal_population}
\end{equation}
where $N$ denotes the set of evaluated agent population sizes. Thus, $n^*$ identifies the configuration with the lowest average token consumption per agent among the evaluated population sizes. It represents an empirical optimum under the selected metric and evaluated population range, rather than a theoretical optimum over all possible agent population sizes.

The following sections empirically examine these scaling characteristics across a diverse set of Web application development tasks.

\subsection{Scaling Behavior and Optimal Agent Population}
\label{sec:optimal_n}

\begin{figure*}[!t]
\centering
\includegraphics[width=\textwidth]{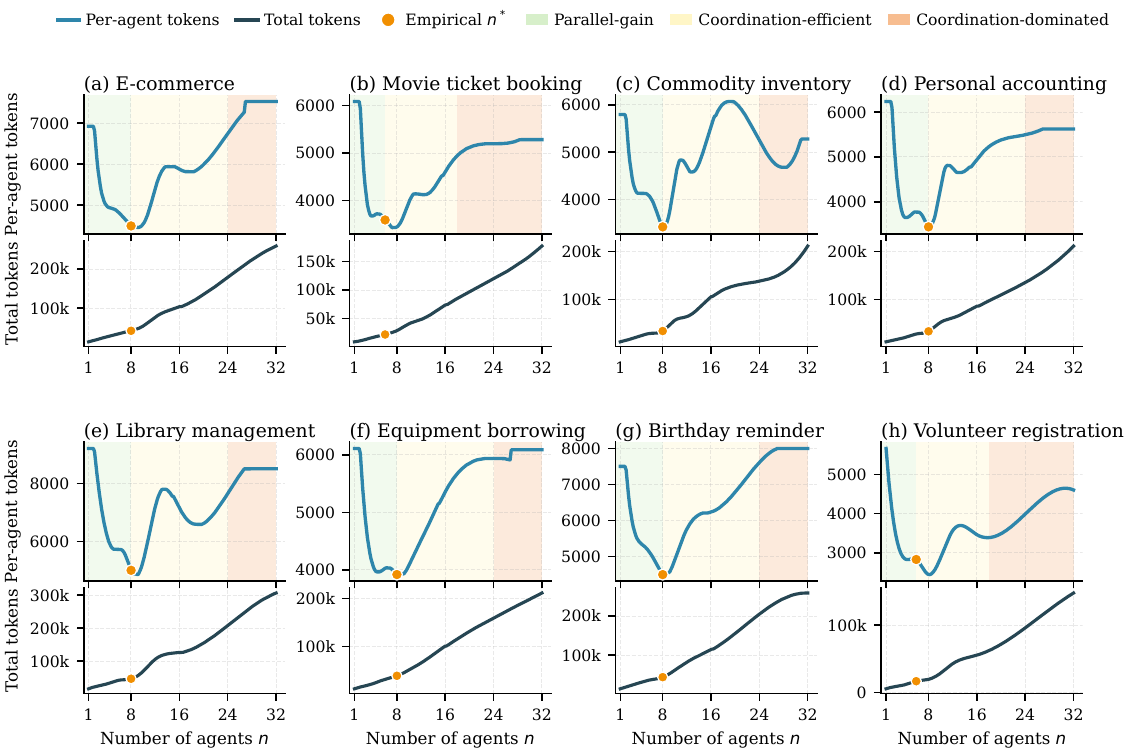}
\caption{Per-agent token consumption (top of each panel) and total token consumption (bottom of each panel) under different agent populations. Panels: (a)~e-commerce, (b)~movie-ticket booking, (c)~commodity inventory, (d)~personal accounting, (e)~library management, (f)~equipment borrowing, (g)~birthday reminder, (h)~volunteer registration. Shaded bands on the per-agent curve mark the parallel-gain, coordination-efficient, and coordination-dominated regimes, and the marker indicates the empirical optimum $n^*$. One legend is shared across panels.}
\Description{Eight task panels. Each panel has a U-shaped per-agent token curve on top and a rising total token curve on the bottom as the agent population increases, with a marker at the empirical optimum.}
\label{fig:scaling_main}
\label{fig:e_commerce}
\label{fig:movie_ticket_booking_per_engineer}
\label{fig:commodity_per_enginner}
\label{fig:personal_accounting_per}
\label{fig:lib_management_u}
\label{fig:equipment_borrowing_u}
\label{fig:birthday_reminder_u}
\label{fig:volunteer_registration_u}
\end{figure*}

We evaluate the population scaling model across a diverse set of application development tasks spanning four categories: high concurrency, data processing, system services, and niche systems. The workflow uses fixed functional roles, including Product Manager (PM), System Architect (SA), Full-Stack Engineer (Eng), and Project Manager (PMgr), while only the number of agents is varied. The evaluated agent populations are $N = \{1,2,4,6,8,10,12,16,24,32\}$. For each task and population size $n$, we record the total token consumption $T_t(n)$ and compute the corresponding per-agent token consumption $T_p(n)$. The empirical optimal population $n^*$ is then identified according to the criterion defined in Section~\ref{sec:scaling_model}. All other task descriptions, role prompts, and decoding parameters are kept fixed, with a temperature of 0.6 using the Moonshot Kimi API (\texttt{kimi-k2-turbo-preview}).

Figure~\ref{fig:scaling_main} reports token consumption for eight Web application tasks spanning high-concurrency, data-processing, system-service, and niche applications. Each panel shows smoothed per-agent token consumption $T_p(n)$ (top) and total token consumption $T_t(n)$ (bottom) for one task. Across tasks, $T_p(n)$ decreases at smaller population sizes and increases after an efficient region, while $T_t(n)$ continues to increase with the agent population. A larger population can therefore reduce average token consumption per agent while increasing overall token expenditure. The empirical optimum is $n^*=8$ for e-commerce, commodity inventory, personal accounting, library management, and equipment borrowing, $n^*=6$ for movie-ticket booking, and $n^*\approx 7.9$ and $n^*\approx 8.2$ for the birthday-reminder and volunteer-registration tasks. On the e-commerce task, the smoothed per-agent curve reaches its minimum near $n=8.2$; we use the empirical value $8$ because $n^*$ in Equation~\eqref{eq:optimal_population} is defined over the discrete set $N$. These optima are concentrated around $n^*\approx 8$. Cross-framework curves aggregated over Django, Gin, and Express are reported in Appendix~\ref{sec:additional_results}.

\subsection{Three-Regime Scaling Behavior}
\label{sec:three_regime}

Across the evaluated tasks, we observe a consistent three-regime scaling pattern as the agent population increases. Rather than improving monotonically with additional agents, token-based efficiency first improves, reaches an efficient region, and then deteriorates as the cost associated with larger agent populations becomes increasingly prominent.

\textbf{Parallel-gain regime.}
When the agent population is small, adding engineering agents reduces per-agent token consumption. The additional agents increase parallel execution capacity, while the cost associated with the larger population remains relatively limited. As a result, the reduction in average token consumption outweighs the additional token expenditure.

\textbf{Coordination-efficient regime.}
At intermediate population sizes, the benefits of additional parallelism and the cost associated with a larger population become relatively balanced. Per-agent token consumption reaches its minimum in this region, with the empirical optima concentrated around $n^*\approx 8$ across the evaluated application development tasks. This regime represents an efficient operating range for the evaluated multi-agent workflow.

\textbf{Coordination-dominated regime.}
When the agent population increases beyond the efficient region, additional agents provide progressively smaller gains in per-agent efficiency, while total token consumption continues to increase. Consequently, per-agent token consumption also increases, indicating that the additional cost of a larger population is no longer offset by improvements in average token efficiency.

Overall, the three regimes characterize a non-linear scaling behavior in agentic Web systems. As the population grows beyond the efficient region, coordination-related costs become increasingly important relative to the benefits of additional agents. These findings highlight that effective scaling depends not only on the number of agents, but also on how their interactions are organized.

\section{From Scaling to Structuring: The Surgical Team Model (STM)}

Motivated by the coordination bottleneck observed in the population scaling analysis, we propose the \textbf{Surgical Team Model (STM)}, a structured coordination framework that organizes agents through role specialization, centralized coordination, and constrained communication. Rather than targeting a fixed agent population size, STM focuses on structuring agent interactions to address coordination challenges as the system scales. This section first introduces the STM architecture and its structured interaction topology, then analyzes the resulting coordination complexity and its scaling implications, and finally evaluates STM through empirical benchmarks against representative baselines.

\subsection{Surgical Team Model for Structured Coordination}
\label{sec:stm_framework}

The scaling results in Section~\ref{sec:three_regime} reveal a fundamental limitation of simply increasing the agent population. We refer to this limitation as the coordination bottleneck, where the coordination requirements associated with a growing agent population can increasingly offset the benefits of additional parallelism. This observation motivates a shift from agent scaling to interaction structuring. To address this bottleneck, we introduce the \textbf{Surgical Team Model (STM)}, a structured coordination framework that organizes agents through three complementary design principles: role specialization, centralized coordination, and bounded communication.

The design of STM is inspired by the surgical team organization described by Brooks, in which a small number of specialized roles are coordinated through a central mechanism rather than unrestricted interactions among all participants. Following this principle, STM assigns agents explicit roles according to their functional responsibilities, which may include architecture design, implementation, testing, documentation, and deployment, depending on the development task. This role specialization establishes a clear division of labor, reduces responsibility overlap and redundant work, and enables agents to focus on well-defined subtasks. It also supports parallel execution when task dependencies permit, allowing specialized contributions to be coordinated toward a common development objective.

At the center of STM is a coordinator that manages the overall development process and serves as the primary interface between the development task and specialist agents. Given a software development task, the coordinator analyzes the requirements, decomposes the task into subtasks, and assigns them to appropriate agents based on their roles and the dependencies among subtasks. During execution, it monitors progress, manages task dependencies, and collects intermediate and final outputs for integration into a coherent solution. For dependent subtasks, the coordinator determines the execution order and forwards relevant intermediate results to subsequent agents. For sufficiently independent subtasks, it enables parallel execution. By separating overall process management from individual task execution, the coordinator provides a unified mechanism for orchestrating development activities and integrating their outputs.

\begin{figure*}[htbp]
\centering
\includegraphics[width=0.8\textwidth]{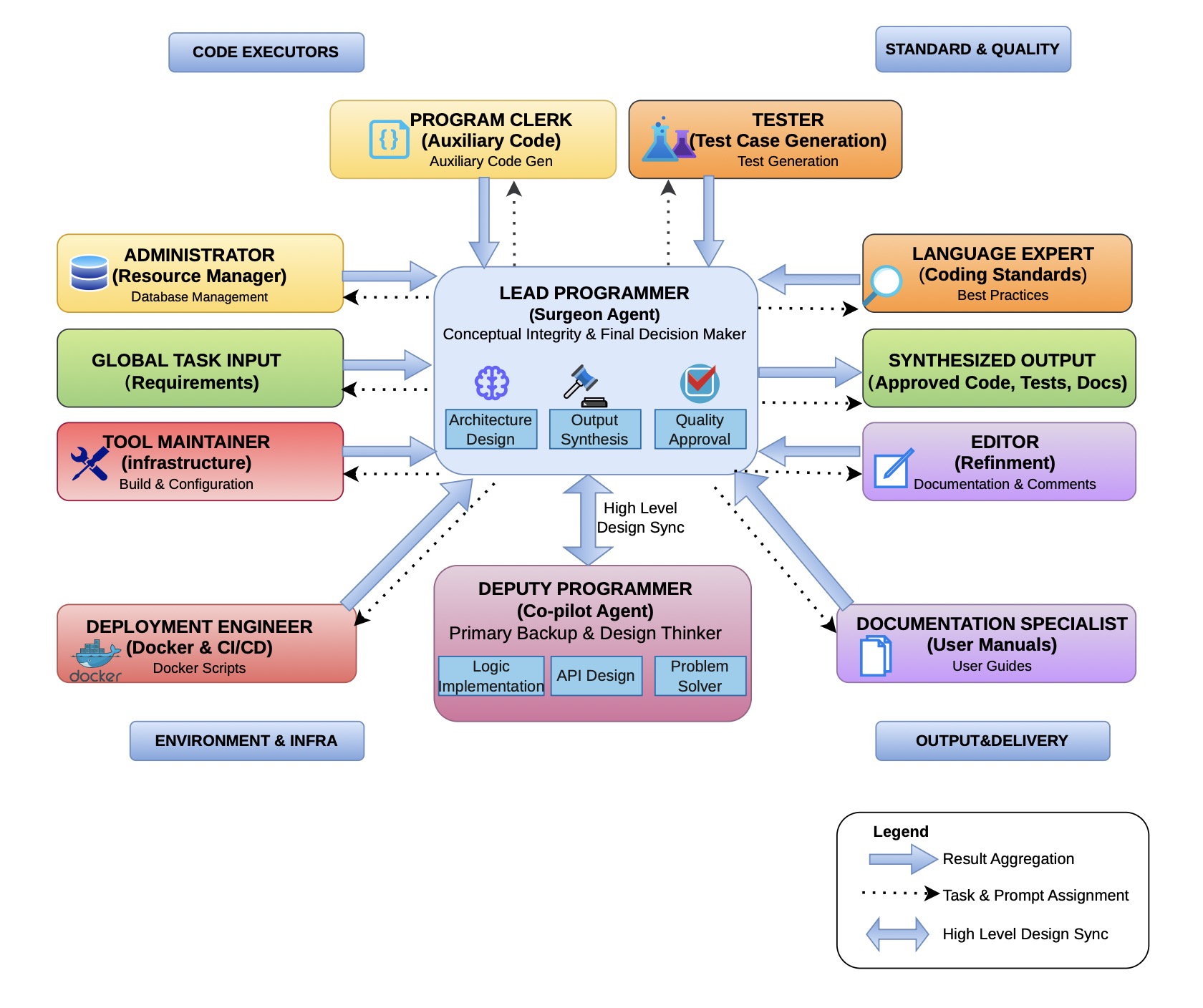}
\caption{\textbf{STM framework with structured interaction topology.}
STM framework with a structured interaction topology. The system adopts a hub-and-spoke architecture in which a central coordinator manages interactions with specialized agents, replacing unrestricted pairwise communication with controlled, coordinator-mediated coordination.}
\Description{Hub-and-spoke diagram of the Surgical Team Model: a central coordinator assigns tasks to and aggregates results from specialized agents such as architect, engineer, tester, and documenter.}
\label{stm}
\end{figure*}

As illustrated in Figure~\ref{stm}, STM organizes interactions between the central coordinator and role-specialized agents through explicit coordination pathways. The topology distinguishes task assignment, result aggregation, and high-level design synchronization, clarifying how information flows between the coordinator and specialists during development. Direct communication between specialists is restricted to interactions required by specific development dependencies, while other coordination is mediated by the central coordinator. This explicit interaction structure provides a basis for controlling communication as the agent population grows.

Together, role specialization, centralized coordination, and bounded communication define the structural principles underlying STM. These principles establish a structured alternative to unrestricted multi-agent collaboration. Building on this design, the following section analyzes the implications of interaction organization for coordination overhead and scalability.

\subsection{Interaction Topology of Structured Coordination}
\label{coordination_modeling}

The scalability implications of STM can be analyzed by examining how the number of coordination relationships grows with the agent population. We compare the interaction complexity of unstructured multi-agent workflows with that of STM, focusing on the growth of potential pairwise communication relationships under different interaction topologies. We model the coordination structure of a multi-agent system with $n$ agents according to the number of inter-agent communication relationships induced by its interaction topology.

\textbf{Unstructured coordination.}
In an unstructured multi-agent workflow, agents may communicate with multiple other agents during task execution. As the agent population increases, the number of potential communication relationships can grow rapidly. In a dense interaction setting, the number of possible pairwise relationships is proportional to $n(n-1)/2$, giving a quadratic upper bound on the interaction space:

$$
C_{\mathrm{unstructured}}(n)=\mathcal{O}(n^2).
$$

This quadratic growth characterizes the potential interaction space rather than the actual number of communication events. It nevertheless indicates that unrestricted interaction can introduce increasingly complex coordination relationships as more agents participate in the development process.

\textbf{STM structured coordination.}
In contrast, STM restricts communication through role specialization and a centralized coordination interface. Each specialist agent communicates through a bounded number of interfaces independently of the total agent population. As a result, the number of communication relationships induced by the STM topology grows approximately linearly with the agent population:

$$
C_{\mathrm{STM}}(n)=\mathcal{O}(n).
$$

Under an unstructured topology, the number of potential communication relationships associated with an agent can increase with the population size. In STM, specialist agents primarily communicate with the central coordinator, keeping their communication degree bounded. Thus, increasing the number of specialist agents does not require a corresponding increase in direct peer-to-peer communication relationships.

\textbf{Scaling implication.}
The difference between these interaction structures provides a structural explanation for their different scaling behavior. As the agent population increases, an unstructured topology may introduce an increasingly large interaction space, whereas STM constrains the growth of topology-induced coordination complexity. This allows additional agents to contribute execution capacity without proportionally increasing the coordination relationships that must be managed by each specialist.

This analysis should be distinguished from the total execution cost of a multi-agent system. The expressions $C_{\mathrm{unstructured}}(n)=\mathcal{O}(n^2)$ and $C_{\mathrm{STM}}(n)=\mathcal{O}(n)$ describe the growth of the interaction structure rather than the complete system cost. In practice, model inference, task complexity, output generation, tool invocation, execution dynamics, and coordination rounds also contribute to the actual cost.

\textbf{Efficiency Model.}
To conceptually characterize the trade-off between task quality and overall system cost, we define collaboration efficiency as:

$$
E(n)=\frac{Q(n)}{C(n)},
$$
where $Q(n)$ represents the quality of the completed task and $C(n)$ denotes the overall system cost associated with the corresponding configuration. This formulation provides a conceptual basis for interpreting the quality--cost trade-off. In the empirical evaluation, however, task quality, token consumption, and execution latency are reported separately rather than combined into a single efficiency score.

\textbf{Empirical Connection.}
In our experiments, the exact coordination cost cannot be directly isolated because agent execution and inter-agent communication are jointly reflected in model token usage. We therefore use total token consumption $T_{\text{total}}(n)$ as an empirical proxy for overall system cost:

$$
C(n)\approx T_{\text{total}}(n).
$$

This proxy captures both execution- and coordination-related token usage and should not be interpreted as a direct measurement of communication cost alone. We use it to compare the overall cost of different agent configurations under the same task and model settings.

The empirical optimal agent population $n^*$ is determined according to the criterion defined in Section~\ref{optimal_size_identification}. This definition identifies the population size with the lowest average token cost per agent among the evaluated configurations. It provides an empirical token-based optimum for the evaluated population sizes and is distinct from the conceptual efficiency formulation $E(n)$.

Overall, the topology analysis and empirical measurements provide complementary views of scalable coordination. The topology characterizes how potential communication relationships grow with the agent population, while the token-based measurements capture the resulting overall system cost during practical execution. Together, they provide a structural and empirical basis for understanding how STM addresses the coordination bottleneck identified in Section~\ref{sec:three_regime}.

\subsection{Empirical Evaluation}
\label{sec:empirical_evaluation}

To empirically evaluate the scaling and coordination model in Section~\ref{coordination_modeling}, we evaluate STM on five full-stack Web application development tasks, including e-commerce, inventory management, equipment borrowing, birthday reminder, and commodity inventory management. All tasks are implemented on a unified technology stack, consisting of a Gin backend, Preact frontend, and PostgreSQL database, to ensure comparability across different systems. Experiments are conducted using the moonshot-v1-8k model via the Kimi API under identical task specifications across all baselines.

\textbf{Experimental Setup.}

The evaluation considers STM and two representative baselines, MetaGPT and Claude Code. MetaGPT represents a role-based multi-agent framework that organizes development through multiple specialized agents, while Claude Code serves as a representative coding-agent baseline. For STM, all experiments follow the coordination structure described in Section~\ref{sec:stm_framework}, where a central coordinator organizes interactions among specialist agents. These baselines provide reference points for evaluating STM against both multi-agent and single-agent development paradigms.

\begin{table*}[t]
    \centering
    \caption{Comprehensive Benchmark: STM Performance VS. MetaGPT}
    \label{tab:comprehensive_metagpt}
    \small 
    \setlength{\tabcolsep}{3.5pt} 
    \begin{tabular}{l|cc|cc|cc|cc|cc}
        \toprule
        \textbf{Task} 
        & \multicolumn{2}{c|}{\textbf{$R_{\text{overall}}$ ($\uparrow$)}} 
        & \multicolumn{2}{c|}{\textbf{$N_{\text{calls}}$ ($\downarrow$)}} 
        & \multicolumn{2}{c|}{\textbf{$L_{\text{eff}}$ ($\uparrow$)}} 
        & \multicolumn{2}{c|}{\textbf{$T_{\text{total}}$ ($\downarrow$)}} 
        & \multicolumn{2}{c}{\textbf{$T_{\text{exec}}$ ($\downarrow$)}} \\
        \cmidrule(lr){2-3} \cmidrule(lr){4-5} \cmidrule(lr){6-7} \cmidrule(lr){8-9} \cmidrule(lr){10-11}
        & Base. & \textbf{STM} & Base. & \textbf{STM} & Base. & \textbf{STM} & Base. & \textbf{STM} & Base. & \textbf{STM} \\
        \midrule
        Birthday Reminder & 65\% & \textbf{86\%} & 15 & \textbf{10} & 174 & \textbf{561} & 2273 & \textbf{8409} & 24.57s & \textbf{128.1s} \\
        Equip. Borrowing & 53\% & \textbf{78\%} & 15 & \textbf{10} & 188 & \textbf{353} & 2192 & \textbf{7348} & 29.29s & \textbf{131.1s} \\
        E-Commerce System & 49\% & \textbf{78\%} & 15 & \textbf{10} & 114 & \textbf{268} & 1664 & \textbf{8058} & 26.79s & \textbf{172.4s} \\
        Inventory Mgmt. & 43\% & \textbf{88\%} & 15 & \textbf{10} & 155 & \textbf{302} & 1745 & \textbf{6321} & 21.69s & \textbf{114.3s} \\
        Commodity Inv. & 43\% & \textbf{88\%} & 15 & \textbf{10} & 83 & \textbf{447} & 1607 & \textbf{8049} & 20.35s & \textbf{134.0s} \\
        \midrule
        \textbf{Avg.} & 51\% & \textbf{84\%} & 15 & \textbf{10} & 143 & \textbf{386} & 1896 & \textbf{7637} & 23.54s & \textbf{136.0s} \\
        \bottomrule
    \end{tabular}
\end{table*}

\begin{table*}[t]
    \centering
    \small
    \caption{Comprehensive Benchmark: STM Performance VS. Claude Code}
    \label{tab:comprehensive_claude}
    \setlength{\tabcolsep}{3.2pt} 
    \begin{tabular}{l|cc|cc|cc|cc|cc}
        \toprule
        \textbf{Task} 
        & \multicolumn{2}{c|}{\textbf{$R_{\text{overall}}$ ($\uparrow$)}} 
        & \multicolumn{2}{c|}{\textbf{$N_{\text{calls}}$ ($\downarrow$)}} 
        & \multicolumn{2}{c|}{\textbf{$L_{\text{eff}}$ ($\uparrow$)}} 
        & \multicolumn{2}{c|}{\textbf{$T_{\text{total}}$ ($\downarrow$)}} 
        & \multicolumn{2}{c}{\textbf{$T_{\text{exec}}$ ($\downarrow$)}} \\
        \cmidrule(lr){2-3} \cmidrule(lr){4-5} \cmidrule(lr){6-7} \cmidrule(lr){8-9} \cmidrule(lr){10-11}
        & Base. & \textbf{STM} & Base. & \textbf{STM} & Base. & \textbf{STM} & Base. & \textbf{STM} & Base. & \textbf{STM} \\
        \midrule
        Birthday Reminder & 71.0\% & \textbf{80.0\%} & 7 & 7 & \textbf{473} & 463 & 6641 & \textbf{6396} & 126.0s & \textbf{124.5s} \\
        Equip. Borrowing  & 92.5\% & \textbf{100.0\%} & 7 & 7 & 558 & \textbf{626} & \textbf{7238} & 7553 & \textbf{136.7s} & 146.7s \\
        E-Commerce System & 87.1\% & 87.1\% & 8 & 8 & 645 & \textbf{656} & \textbf{8699} & 9260 & \textbf{161.2s} & 171.8s \\
        Inventory Mgmt.   & 84.8\% & \textbf{87.2\%} & 8 & 8 & \textbf{1070} & 861 & 1174 & 1174 & \textbf{225.3s} & 462.8s \\
        Commodity Inv.    & \textbf{74.6\%} & 74.3\% & 8 & 8 & \textbf{1078} & 878 & \textbf{12650} & 12802 & \textbf{149.3s} & 166.7s \\
        \midrule
        \textbf{Avg.} & 82.0\% & \textbf{85.7\%} & 7.6 & 7.6 & \textbf{764.8} & 696.8 & \textbf{7280.4} & 7437.0 & \textbf{159.7s} & 214.5s \\
        \bottomrule
    \end{tabular}
\end{table*}

\subsubsection{Evaluation Metrics}

We evaluate the systems from the following metrics. 

\textbf{Cost Approximation.}
Following the coordination model in Section~\ref{coordination_modeling}, we use total token consumption as an empirical proxy for the overall system cost. Specifically, $C(n)\approx T_{\text{total}}(n)$ is used as an empirical approximation, where $T_{\text{total}}(n)$ denotes the total number of input and output tokens consumed by the system with $n$ agents. This proxy captures both inter-agent coordination and execution-related token usage. It therefore reflects the overall token-level cost of the system rather than serving as a direct measurement of communication cost alone.

\textbf{Quality Definition.}
We measure task quality using effective code lines $L_{\text{eff}}$ and the overall completion rate $R_{\text{overall}}$. The effective code lines represent the number of functional lines of generated code that contribute to successful task completion, providing a quantitative measure of the amount of usable implementation produced by each system. The overall completion rate is defined as $R_{\text{overall}} = \sum_{i=1}^{m} R_i w_i$, where $R_i \in [0,1]$ denotes the completion rate of development task $i$, $m$ is the number of evaluated tasks, and $w_i$ is the weight assigned to task $i$, with $\sum_{i=1}^{m} w_i = 1$. For each task, $R_i$ reflects the proportion of required functionality successfully implemented and validated. When all tasks are weighted equally, $w_i = 1/m$. This metric measures the extent to which the generated applications satisfy their required functionality, while $L_{\text{eff}}$ provides a complementary measure of the functional implementation produced by each system.

\textbf{System Efficiency.}
We use $E(n)=\frac{Q(n)}{C(n)}$ as a conceptual formulation of the quality--cost trade-off introduced in Section~\ref{coordination_modeling}. Here, $Q(n)$ represents task quality and $C(n)$ represents the overall system cost. The formulation provides a conceptual basis for interpreting whether improvements in task quality are obtained at a reasonable system cost. In the empirical evaluation, however, we report task quality, token consumption, and execution latency separately rather than combining them into a single efficiency score. This separation avoids introducing an additional weighting scheme among heterogeneous evaluation metrics.

\textbf{Operational Metrics.}
We additionally track implementation-level statistics to characterize the execution behavior of different systems. Total token consumption is defined as $T_{\text{total}} = T_{\text{prompt}} + T_{\text{completion}}$, where $T_{\text{prompt}}$ and $T_{\text{completion}}$ denote input and output token usage, respectively. Execution time is measured as $T_{\text{exec}} = T_{\text{end}} - T_{\text{start}}$, where $T_{\text{start}}$ and $T_{\text{end}}$ denote the beginning and end timestamps of task execution. We also report the number of agent calls $N_{\text{calls}}$ to characterize the interaction overhead associated with each system. These metrics provide complementary views of system cost: token consumption reflects the amount of model computation and generated information, agent calls characterize interaction frequency, and execution time reflects the end-to-end latency of completing a development task.

\subsubsection{Benchmark Comparison}

We compare STM against MetaGPT and Claude Code in terms of task completion quality, token-based system cost, and implementation-level operational metrics, as shown in Table~\ref{tab:comprehensive_metagpt} and Table~\ref{tab:comprehensive_claude}. The two comparisons provide complementary reference points for evaluating the practical effects of STM's structured coordination design.

\textbf{Task Completion Quality.} STM significantly outperforms MetaGPT in task completion quality, achieving an average completion rate of 84\% compared with 51\%. It also produces approximately 2.7$\times$ more effective code lines ($L_{\text{eff}}$), with average values of 386 and 143, respectively. These results indicate that STM produces more functional implementation under the evaluated development tasks.

Compared with Claude Code, STM achieves an average completion rate of 85.7\%, while Claude Code achieves 82.0\%. The difference is relatively small, indicating that STM remains competitive with a strong coding-agent baseline despite employing a multi-agent coordination structure. The results therefore suggest that introducing multiple specialized agents does not necessarily compromise task completion quality when their interactions are organized through a constrained communication structure.

\textbf{System Cost and Operational Metrics.} STM exhibits a different cost profile from the two baselines. Compared with MetaGPT, STM uses fewer agent calls on average, with 10 calls compared with 15 for MetaGPT, but consumes more total tokens, with 7637 tokens compared with 1896 tokens. This result indicates that reducing the number of agent interactions does not necessarily reduce overall token consumption. The structured workflow may instead involve more substantial generation within each interaction, particularly when specialist agents are responsible for producing more complete outputs within their assigned roles.

The execution time also differs substantially between the two systems. STM requires an average of 136.0 seconds, compared with 23.54 seconds for MetaGPT. Thus, although STM reduces the number of agent calls, its individual interactions are more computationally substantial and result in higher end-to-end execution time. This observation is consistent with the distinction made in Section~\ref{coordination_modeling} between topology-level coordination complexity and total execution cost.

Compared with Claude Code, STM has comparable average total token consumption, with 7437 tokens for STM and 7280.4 tokens for Claude Code. Both systems also use the same average number of agent calls, 7.6. However, STM requires more execution time on average, with 214.5 seconds compared with 159.7 seconds for Claude Code. These results indicate that STM maintains a comparable token-level cost while introducing additional execution latency in the evaluated setting.

\textbf{Efficiency.} The results demonstrate a clear quality--cost trade-off for STM. Compared with MetaGPT, STM substantially improves task completion quality and effective code generation, at the expense of higher token consumption and execution time. In other words, the additional system cost is accompanied by a substantial improvement in functional output. This trade-off is consistent with the conceptual efficiency formulation $E(n)=Q(n)/C(n)$, but we do not reduce the empirical results to a single efficiency score.

Compared with Claude Code, STM achieves a slightly higher average completion rate, with 85.7\% compared with 82.0\%, while maintaining comparable total token consumption. The difference in completion quality is relatively small, suggesting that STM can provide competitive task performance without substantially increasing token-level system cost. However, the higher execution time indicates that token-level cost and end-to-end latency capture different aspects of system efficiency.

Overall, STM achieves competitive task completion quality while maintaining comparable token consumption to Claude Code, although with higher execution latency. These results suggest that structured coordination can support scalable multi-agent development by organizing agent interactions without requiring a fully decentralized interaction structure.

\textbf{Scalability and Discussion.} The results highlight the practical trade-offs associated with STM's structured coordination design. Compared with MetaGPT, STM achieves higher task completion rates and produces more effective code lines while requiring fewer agent calls. However, these improvements come with higher token consumption and execution latency. Compared with Claude Code, STM achieves competitive task completion quality and comparable token consumption, but incurs higher execution latency. These findings indicate that the benefits of structured coordination are reflected primarily in task completion quality and the organization of agent interactions, rather than in uniformly lower execution costs.

These empirical observations complement the topology-level analysis in Section~\ref{coordination_modeling}. The theoretical analysis characterizes the growth of potential communication relationships under different interaction topologies, whereas the benchmark results measure overall token consumption and end-to-end execution latency. The latter are also affected by model inference, task complexity, output generation, and execution dynamics. Therefore, the linear growth of STM's topology-induced coordination relationships should not be interpreted as evidence of linear growth in total execution cost or as a guarantee of lower empirical cost. Instead, the topology-level analysis explains the structural scalability of STM, while the empirical results characterize its practical performance under different quality, token-consumption, and execution-latency requirements.

\vspace{-1.2em}

\section{Conclusion}

This work investigates agent population scaling in agentic Web systems and identifies a non-linear pattern in which per-agent token consumption initially decreases but eventually increases as coordination-related costs grow. Motivated by this finding, we propose STM, a structured coordination framework based on specialized roles and bounded interaction. By organizing collaboration through a central coordinator, STM provides a systematic way to manage task decomposition, information exchange, and result integration as agent populations grow. Experiments on full-stack Web application development tasks show that STM achieves competitive task completion quality with token consumption comparable to that of a strong coding-agent baseline. These findings highlight the importance of structured coordination in designing scalable multi-agent development systems and demonstrate the need to consider interaction organization alongside agent population size.

Future work will explore adaptive coordination mechanisms, including dynamic role allocation and communication strategies that respond to task complexity and resource constraints. We will also investigate cross-model collaboration and evaluate these approaches across diverse Web application development scenarios to further improve the scalability and robustness of agentic Web systems.

\section*{Ethics and Privacy Statement}
This work does not involve human participants or personal data, and therefore does not raise issues related to informed consent or human-subject privacy. The proposed structured coordination approach is intended to improve the scalability and efficiency of multi-agent software development; potential risks mainly concern the future misuse of autonomous agent systems, such as unintended software generation or deployment. We encourage the use of appropriate human oversight and safety mechanisms when deploying such systems in real-world Web environments.

\bibliographystyle{ACM-Reference-Format}
\bibliography{software}

@Book{brooks1995mythical,
  author =       "Frederick P. Brooks Jr.",
  title =        "The Mythical Man-Month: Essays on Software Engineering",
  publisher =    "Addison-Wesley",
  year =         1995,
  address =      "Boston, MA",
  edition =      "2nd"
}

@misc{wang2023metagpt,
  author    = {Hong, Sirui and
               Zhuge, Mingchen and
               Chen, Jiaqi and
               Zheng, Xiawu and
               Cheng, Yuheng and
               Zhang, Ceyao and
               Wang, Jinlin and
               Wang, Zili and
               Yau, Steven Ka Shing and
               Lin, Zijuan and
               Zhou, Liyang and
               Ran, Chenyu and
               Xiao, Lingfeng and
               Wu, Chenglin and
               Schmidhuber, J{\"u}rgen},
  title     = {MetaGPT: Meta Programming for A Multi-Agent Collaborative Framework},
  year      = {2023},
  archivePrefix = {arXiv},
  eprint    = {2308.00352},
  doi       = {10.48550/arXiv.2308.00352}
}

@Article{zhu2024sparse,
  author        = {Chen, Dong and
                   Zhang, Kaixiang and
                   Wang, Yongqiang and
                   Yin, Xunyuan and
                   Li, Zhaojian and
                   Filev, Dimitar},
  title         = {Communication-Efficient Decentralized Multi-Agent Reinforcement Learning for Cooperative Adaptive Cruise Control},
  journal       = {IEEE Transactions on Intelligent Vehicles},
  volume        = {9},
  number        = {10},
  pages         = {6436--6449},
  month         = {Oct.},
  year          = {2024},
  publisher     = {IEEE},
  doi           = {10.1109/TIV.2024.3368025},
}

@Article{li2023adaptive,
  author        = {Wang, Jianrui and
                   Li, Yi and
                   Hong, Yitian and
                   Tang, Yang},
  title         = {Integrated Adaptive Communication in Multi-Agent Systems: Dynamic Topology, Frequency, and Content Optimization for Efficient Collaboration},
  journal       = {Neurocomputing},
  volume        = {613},
  pages         = {129068},
  month         = {Feb.},
  year          = {2025},
  publisher     = {Elsevier},
  doi           = {10.1016/j.neucom.2024.129068},
}

@inproceedings{qian2025macnet,
  title        = {Scaling Large Language Model-based Multi-Agent Collaboration},
  author       = {Qian, Chen and
                  Xie, Zihao and
                  Wang, YiFei and
                  Liu, Wei and
                  Zhu, Kunlun and
                  Xia, Hanchen and
                  Dang, Yufan and
                  Du, Zhuoyun and
                  Chen, Weize and
                  Yang, Cheng and
                  Liu, Zhiyuan and
                  Sun, Maosong},
  booktitle    = {International Conference on Learning Representations},
  year         = 2025,
 pages = {41488--41505},
address      = {Singapore},
  publisher    = {OpenReview.net},

}

@inproceedings{du2023chatdev,
  title        = {ChatDev: Communicative Agents for Software Development},
  author       = {Qian, Chen and
                  Liu, Wei and
                  Liu, Hongzhang and
                  Chen, Nuo and
                  Dang, Yufan and
                  Li, Jiahao and
                  Yang, Cheng and
                  Chen, Weize and
                  Su, Yusheng and
                  Cong, Xin and
                  Xu, Juyuan and
                  Li, Dahai and
                  Liu, Zhiyuan and
                  Sun, Maosong},
  booktitle    = {Proceedings of the 62nd Annual Meeting of the Association for Computational Linguistics},
  pages        = {15174--15186},
  year         ={ 2024},
  publisher    = {Association for Computational Linguistics},
  address      = {Bangkok, Thailand},
}

@misc{jiang2024crewai,
  author    = {Duan, Zhihua and  Wang, Jialin},
  title     = {Exploration of LLM Multi-Agent Application Implementation Based on LangGraph+CrewAI},
  year      = {2024},
  archivePrefix = {arXiv},
  eprint    = {2411.18241},
  doi       = {10.48550/arXiv.2411.18241}
}

@inproceedings{jiang2023autogen,
  title        = {AutoGen: Enabling Next-Gen LLM Applications via Multi-Agent Conversation},
  author       = {Wu, Qingyun and
                  Bansal, Gagan and
                  Zhang, Jieyu and
                  Wu, Yiran and
                  Li, Beibin and
                  Zhu, Erkang and
                  Jiang, Li and
                  Zhang, Xiaoyun and
                  Zhang, Shaokun and
                  Liu, Jiale and
                  Awadallah, Ahmed Hassan and
                  White, Ryen W. and
                  Burger, Doug and
                  Wang, Chi},
  booktitle    = {Proceedings of the 1st Conference on Language Modeling},
  year         = {2024},
  pages        = {1--46},
  address      = {Philadelphia, PA, USA},
}

@inproceedings{chen2023agentverse,
  title        = {AgentVerse: Facilitating Multi-Agent Collaboration and Exploring Emergent Behaviors},
  author       = {Chen, Weize and
                  Su, Yusheng and
                  Zuo, Jingwei and
                  Yang, Cheng and
                  Yuan, Chenfei and
                  Chan, Chi-Min and
                  Yu, Heyang and
                  Lu, Yaxi and
                  Hung, Yi-Hsin and
                  Qian, Chen and
                  Qin, Yujia and
                  Cong, Xin and
                  Xie, Ruobing and
                  Liu, Zhiyuan and
                  Sun, Maosong and
                  Zhou, Jie},
  booktitle    = {International Conference on Learning Representations},
  pages        = {20094--20136},
  year         = 2024,
  publisher    = {OpenReview.net},
  address      = {Vienna, Austria},
}

@inproceedings{paranjape2024codeagents,
  author    = {Paranjape, Aditya and Morris, John X. and Hofmann, Valentin and others},
  title     = {CodeAgents: Evaluating LLMs with Communicative Agents for Code Review},
  booktitle = {Proceedings of EMNLP},
  pages     = {10987--11002},
  year      = {2024}, 
address={Miami, Florida, USA},
 publisher   = {Association for Computational Linguistics},
}

@inproceedings{li2023camel,
  title        = {CAMEL: Communicative Agents for "Mind" Exploration of Large Language Model Society},
  author       = {Li, Guohao and
                  Hammoud, Hasan Abed Al Kader and
                  Itani, Hani and
                  Khizbullin, Dmitrii and
                  Ghanem, Bernard},
  booktitle    = {Advances in Neural Information Processing Systems 36},
  pages        = {51991--52008},
  year         = {2023},
  publisher    = {Neural Information Processing Systems Foundation, Inc.},
  address      = {New Orleans, Louisiana, USA},
}

@misc{google2025scaling,
  author    = {Kim, Yubin and
               Gu, Ken and
               Park, Chanwoo and
               Park, Chunjong and
               Schmidgall, Samuel and
               Heydari, A. Ali and
               Yan, Yao and
               Zhang, Zhihan and
               Zhuang, Yuchen and
               Liu, Yun and
               Malhotra, Mark and
               Liang, Paul Pu and
               Park, Hae Won and
               Yang, Yuzhe and
               Xu, Xuhai and
               Du, Yilun and
               Patel, Shwetak and
               Althoff, Tim and
               McDuff, Daniel and
               Liu, Xin},
  title     = {Towards a Science of Scaling Agent Systems},
  year      = {2025},
  archivePrefix = {arXiv},
  eprint    = {2512.08296},
  doi       = {10.48550/arXiv.2512.08296}
}

@misc{li2024moreagents,
  author    = {Li, Junyou and
               Zhang, Qin and
               Yu, Yangbin and
               Fu, Qiang and
               Ye, Deheng},
  title     = {More Agents Is All You Need},
  year      = {2024},
  archivePrefix = {arXiv},
  eprint    = {2402.05120},
  doi       = {10.48550/arXiv.2402.05120}
}

@misc{tran2026singleagent,
  author    = {Tran, Dat and
               Kiela, Douwe},
  title     = {Single-Agent LLMs Outperform Multi-Agent Systems on Multi-Hop Reasoning Under Equal Thinking Token Budgets},
  year      = {2026},
  archivePrefix = {arXiv},
  eprint    = {2604.02460},
  doi       = {10.48550/arXiv.2604.02460}
}

@inproceedings{zheng2025thought,
  title        = {Thought Communication in Multiagent Collaboration},
  author       = {Zheng, Yujia and
                  Zhao, Zhuokai and
                  Li, Zijian and
                  Xie, Yaqi and
                  Gao, Mingze and
                  Zhang, Lizhu and
                  Zhang, Kun},
  booktitle    = {Advances in Neural Information Processing Systems 38},
  pages        = {136784--136813},
  year         = 2025,
  publisher    = {Curran Associates, Inc.},
  address      = {San Diego, California, USA and Mexico City, Mexico},
}

@Article{kim2026capable,
  author        = {Kim, Yubin and
                   Gu, Ken and
                   Park, Chanwoo and
                   Park, Chunjong and
                   Schmidgall, Samuel and
                   Heydari, A. Ali and
                   Yan, Yao and
                   Zhang, Zhihan and
                   Zhuang, Yuchen and
                   Liu, Yun and
                   Malhotra, Mark and
                   Liang, Paul Pu and
                   Park, Hae Won and
                   Yang, Yuzhe and
                   Xu, Xuhai and
                   Du, Yilun and
                   Patel, Shwetak and
                   Althoff, Tim and
                   McDuff, Daniel and
                   Liu, Xin},
  title         = {Capable Language Models Can Outgrow the Benefits of Collaboration},
  journal       = {Nature Machine Intelligence},
  volume        = {8},
  number        = {7},
  pages         = {1157--1172},
  month         = {July},
  year          = {2026},
  publisher     = {Springer Nature},
  doi           = {10.1038/s42256-026-01268-y},
}

@inproceedings{yang2026understanding,
  title={Understanding Agent Scaling in LLM-Based Multi-Agent Systems via Diversity},
  author={Yang, Yingxuan and Qu, Chengrui and Wen, Muning and Shi, Laixi and Wen, Ying and Zhang, Weinan and Wierman, Adam and Gu, Shangding},
  booktitle={ICLR 2026 Workshop on AI for Mechanism Design and Strategic Decision Making (AIMS)},
  year={2026},
pages={1-26},
publisher = {OpenReview.net},
  address={Singapore}
}

@inproceedings{chen2026diversitycollapse,
  title={Diversity Collapse in Multi-Agent LLM Systems: Structural Coupling and Collective Failure in Open-Ended Idea Generation},
  author={Chen, Nuo and Tong, Yicheng and Yang, Yuzhe and He, Yufei and Zhang, Xueyi and Zou, Qingyun and Wang, Qian and He, Bingsheng},
  booktitle={Findings of the Association for Computational Linguistics: ACL 2026},
  year={2026},
  pages={251--306},
  publisher={Association for Computational Linguistics},
  address={San Diego, California, United States}
}

@inproceedings{Su2026,
  author       = {Jinwei Su and
                  Qizhen Lan and
                  Yinghui Xia and
                  Lifan Sun and
                  Weiyou Tian and
                  Tianyu Shi and
                  Lewei He},
  title        = {Difficulty-Aware Agentic Orchestration for Query-Specific Multi-Agent Workflows},
  booktitle    = {Proceedings of the ACM Web Conference (WWW)},
  year         = {2026},
  pages        = {2060--2070},
  publisher    = {ACM},
  address      = {Dubai, United Arab Emirates}
}

@inproceedings{Dong2026,
  author       = {Guanting Dong and
                  Licheng Bao and
                  Zhongyuan Wang and
                  Kangzhi Zhao and
                  Xiaoxi Li and
                  Jiajie Jin and
                  Jinghan Yang and
                  Hangyu Mao and
                  Fuzheng Zhang and
                  Kun Gai and
                  Guorui Zhou and
                  Yutao Zhu and
                  Ji-Rong Wen and
                  Zhicheng Dou},
  title        = {Toward Generalized Web Agent Training: A Deep Dive into Entropy-Balanced Reinforcement Learning},
  booktitle    = {Proceedings of the ACM Web Conference (WWW)},
  year         = {2026},
  pages        = {2126--2137},
  publisher    = {ACM},
  address      = {Dubai, United Arab Emirates}
}

\FloatBarrier
\appendix

\section{Additional Experimental Results}
\label{sec:additional_results}

\begin{figure*}[t]
\centering
\includegraphics[width=\textwidth]{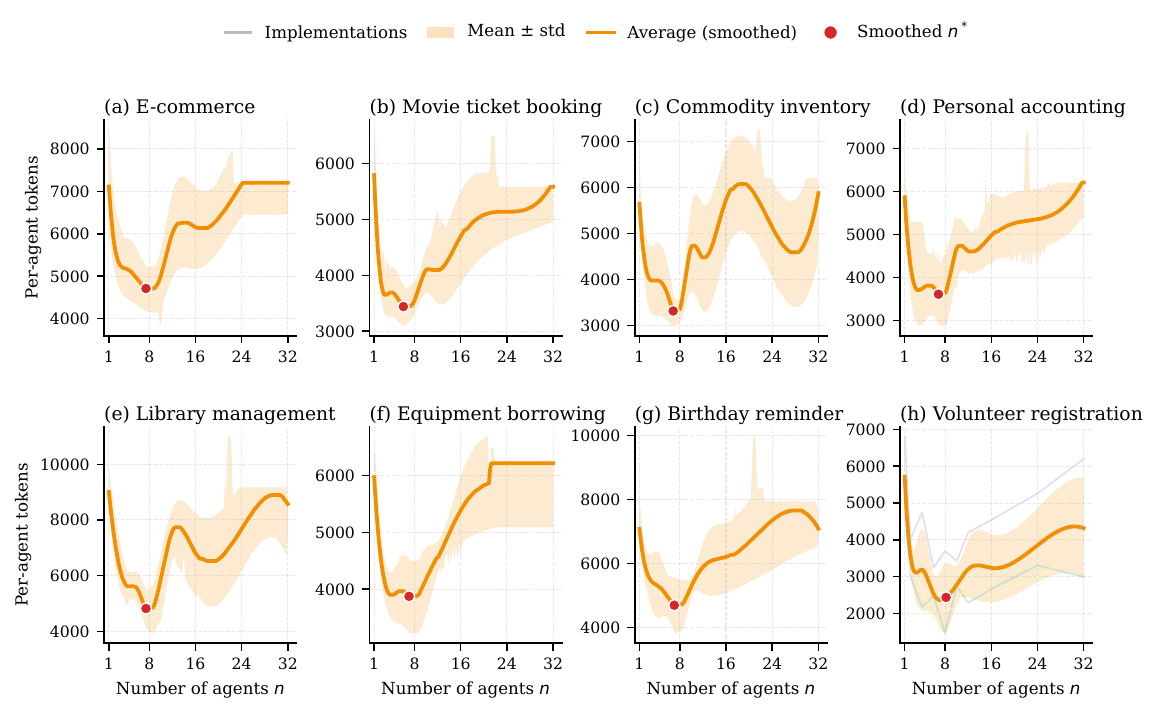}
\caption{Cross-framework per-agent token consumption. Panels: (a)~e-commerce, (b)~movie-ticket booking, (c)~commodity inventory, (d)~personal accounting, (e)~library management, (f)~equipment borrowing, (g)~birthday reminder, (h)~volunteer registration. Each panel aggregates Django, Gin, and Express implementations of one task. The band is the mean $\pm$ standard deviation, and the marker indicates the smoothed optimum. One legend is shared across panels.}
\Description{Eight panels of cross-framework per-agent token curves with shaded standard-deviation bands and a marker at the smoothed optimum for each Web application task.}
\label{fig:scaling_cf}
\label{fig:e_commerce2}
\label{fig:movie_ticket_total}
\label{fig:commodity_total}
\label{fig:personal_accounting_total}
\label{fig:lib_management_total}
\label{fig:equipment_borrowing_total}
\label{fig:lib_birthday_reminder}
\label{fig:volunteer_registration_total}
\end{figure*}

Figure~\ref{fig:scaling_main} in the main text shows per-agent and total token consumption for each task. This appendix examines the same eight tasks through the cross-framework curves in Figure~\ref{fig:scaling_cf}, where each panel aggregates Django, Gin, and Express. Across these settings, per-agent token consumption remains approximately U-shaped, and the optimal team size stays in the same range as in the main experiments.

\subsection{High-Concurrency Systems}

This category includes event-driven systems with bursty requests and strong real-time constraints. These tasks are sensitive to coordination overhead due to frequent inter-agent interactions.

\textbf{Instance: Movie Ticket Booking System.}

As illustrated in Figure~\ref{fig:scaling_main}(b) and Figure~\ref{fig:scaling_cf}(b), the movie ticket booking system exhibits an optimal team size of $n^*=6$. The per-agent curve shows a transition from the \textit{Parallel-Gain Regime} to the \textit{Coordination-Dominated Regime}, with the minimum indicating the empirically optimal team size.

Cross-framework aggregation across Django, Gin, and Express yields $n^*\approx6.9$, indicating consistent scaling behavior across different backend implementations.

\subsection{Data Processing Systems}

This category involves structured workflows such as data collection, cleaning, and storage. Compared with event-driven systems, coordination costs are primarily associated with the structure and dependencies of the processing pipeline.

\textbf{Instance: Commodity Inventory Management Tool.}

As shown in Figure~\ref{fig:scaling_main}(c) and Figure~\ref{fig:scaling_cf}(c), the commodity inventory management system exhibits an optimal team size of $n^*=8$. The U-shaped curve indicates that moderate task decomposition reduces the per-agent token cost, while further scaling introduces increasing coordination overhead.

Cross-framework aggregation across Django, Gin, and Express yields $n^*\approx7.6$, indicating stable scaling behavior across different implementations.

\textbf{Instance: Personal Accounting Applet.}

As shown in Figure~\ref{fig:scaling_main}(d) and Figure~\ref{fig:scaling_cf}(d), the personal accounting applet exhibits an optimal team size of $n^*=8$. The relatively flat U-shaped curve around the minimum suggests a wider range of team sizes with similar per-agent token consumption.

Cross-framework aggregation yields $n^*\approx7.6$, consistent with the scaling behavior observed across the other systems.

\subsection{System Service Tasks}

These tasks feature modular architectures with relatively decoupled components, resulting in comparatively stable coordination patterns.

\textbf{Instance: Library Management System.}

As shown in Figure~\ref{fig:scaling_main}(e) and Figure~\ref{fig:scaling_cf}(e), the library management system exhibits an optimal team size of $n^*=8$. The results indicate that task decomposition remains beneficial at moderate team sizes, while additional agents introduce coordination overhead beyond the optimum.

Cross-framework aggregation across Django, Gin, and Express yields $n^*\approx8.2$, suggesting stable scaling behavior across different backend configurations.

\textbf{Instance: Equipment Borrowing System.}

As shown in Figure~\ref{fig:scaling_main}(f) and Figure~\ref{fig:scaling_cf}(f), the equipment borrowing system exhibits an optimal team size of $n^*=8$. The results show the same non-linear scaling pattern, with increasing coordination overhead as the team size grows beyond the optimum.

\subsection{Niche Applications}

Niche applications involve lightweight business logic and relatively limited concurrency. Despite their simplicity, these systems still exhibit non-linear scaling behavior as the agent population increases.

\textbf{Instance: Birthday Reminder System.}

As shown in Figure~\ref{fig:scaling_main}(g) and Figure~\ref{fig:scaling_cf}(g), the birthday reminder system exhibits an optimal team size of $n^*=7.9$. The results indicate that increasing the agent population beyond the optimum introduces additional coordination overhead without reducing the per-agent token cost.

\textbf{Instance: Volunteer Registration System.}

As shown in Figure~\ref{fig:scaling_main}(h) and Figure~\ref{fig:scaling_cf}(h), the volunteer registration system exhibits an optimal team size of $n^*=8.2$. Despite differences in task characteristics, the same non-linear scaling pattern is observed.

\end{document}